\documentstyle[12pt]{article}

\begin{document}
{\sf \begin{center} \noindent {\Large \bf Plasma fast torus dynamos versus laminar plasma dynamos in laboratory}\\[3mm]

by \\[0.3cm]

{\sl L.C. Garcia de Andrade}\\

\vspace{0.5cm} Departamento de F\'{\i}sica
Te\'orica -- IF -- Universidade do Estado do Rio de Janeiro-UERJ\\[-3mm]
Rua S\~ao Francisco Xavier, 524\\[-3mm]
Cep 20550-003, Maracan\~a, Rio de Janeiro, RJ, Brasil\\[-3mm]
Electronic mail address: garcia@dft.if.uerj.br\\[-3mm]
\vspace{2cm} {\bf Abstract}
\end{center}
\paragraph*{}
Earlier Wang et al [Phys Plasmas (2002)] have estimated a growth
rate for the magnetic field of ${\gamma}=0.055$ and flow ionization
velocity of $51{km}/{s}$ in a laminar plasma slow dynamo mode for
aspect ratio of ${r_{0}}/L\approx{0.6}$, where $r_{0}$ is the
internal straight cylinder radius, and L is the length scale of the
plasma cylinder. In this paper, fast dynamo modes in curved
Riemannian heliotron are shown to be excited on a plasma flow
yielding a growth rate of ${\gamma}=0.318$ for an aspect ratio of
${r_{0}}/L\approx{0.16}$. It is interesting to note that the first
growth rate was obtained in the Wang et al slow dynamo, where the
magnetic Reynolds number of $Re_{m}=210$, while in the second one
considered in this paper one uses the limit of
$Re_{m}\rightarrow{\infty}$. These growth rates ${\gamma}$ are
computed by applying the fast dynamo limit
$lim_{{\eta}\rightarrow{0}}{\gamma}(\eta)>0$. This limit is used in
the self-induced equation, without the need to solve these equations
to investigate the fast dynamo action of the flow. In this sense the
fast dynamo seems to be excited by the elongation of the plasma
device as suggested by Wang group. The Frenet curvature of the tube
is given by ${\kappa}_{0}\approx{0.5 m^{-1}}$. It is suggested that
the small Perm torus could be twisted [Dobler et al, Phys Rev E
(2003)] in order to enhance even more the fast dynamo effect. By
considering the stability of the plasma torus one obtains a value
for the fast dynamo growth rate as high as ${\gamma}=1.712$ from a
general expression ${\gamma}=0.16{\omega}$ and a toroidal
oscillation of a chaotic flow of ${\omega}=\frac{2{\pi}}{6}$.{\bf
PACS numbers:\hfill\parbox[t]{13.5cm}{91.25.Cw-dynamo theories.
02.40.Hw:differential geometries.}}}

\newpage
\newpage
 Though dynamo theory \cite{1,2} has been suffer a boom of development in the last two or four decades. Despite of this fact the experimental framework
 has not been suffer the same deal of success and development and so far few attempts, such as the Perm torus and Riga experiment \cite{3} have
 achieved some success in detecting dynamo action and self-sustaining magnetic fields. More recently , however, Wang and his group \cite{4} has
 developed
 the first flowing magnetic plasma (FMP) experiment, called $P-24$
 \cite{5} to detect dynamo action. In these investigations plasma torus and rotating flows are used within the metalic curved and sometimes twisted
 container, where the magnetized plasma is confined. Plasma investigations were benifitiated in the past by the use of
 Riemannian geometry, where the plasma devices such as heliotrons,
 stellarators and tokamaks could be described by a Riemann metric geometry. Earlier
 studies by Mikhailovskii \cite{6}, have made use of non-diagonal
 Riemann metrics to describe
 such a plasma devices. More recently Ricca \cite{7} has made use of a diagonal twisted magnetic flux tube metric to investigations in plasma astrophysics.
 Yet more recently Ricca's metric has been applied to the investigation of several problems in plasma physics and astrophysics such as curved and twisted
 currents in solar loops \cite{8} as in the almost helical plasmas in electron-magnetohydrodynamics (EMHD) \cite{8}. Earlier the first example of a chaotic
 fast dynamo solution was found by Arnold et al \cite{9,10} by making
 use of a compressed and stretched Riemannian metric of the dynamo
 flow. More recently conformal stretching dynamos have also been
 investigated \cite{11} by using the Vainshtein-Zeldovich stretch-twist and fold (STF) fast dynamo method generation \cite{11}. Recently an interesting type of
 sodium dynamo have been obtained by Wang et al \cite{4} making use of a laminar plasma slow dynamo of aspect ratio of $0.6$ and found a flow velocity
 estimate of $51km/s$. In their paper, they suggested that the
 elongation of the plasma device could excite fast dynamo modes. In
 this work one makes an application of their idea by considering the
 stretching of the Riemannian plasma flux tube. The simple
 stretching process is per se, already a fundamental process on the
 existence of fast dynamo action \cite{2}. Thus in this paper,  one
 considers a Riemannian heliotron which aspect ratio is given by
 $0.001$ which is less than the laminar plasma cylinder one. The
 estimate on the plasma flow velocity used is $50km/s$ around the
 value obtained by Wang et al in order to determine the geometry of the twisted torus or heliotron. A distinct
 feature one considers here with respect to Wang et al work, is
 that the Beltrami flow that is very suitable for fast dynamo
 action, is not considered here and that only incompressible flow
 assumption is made. Frenet curvature is considered too weak and
 that the poloidal component of the magnetic field does not depend
 upon the toroidal coordinate-s. In liquid metals velocities are much lower than the plasma ones and are around $20m/s$. One another interesting aspect is
 that contrary to Zhang et al paper , one does not use any numerical
 code to solve the magnetic self-induction equation, and only the
 fast dynamo limit $lim_{{\eta}\rightarrow{0}}{\gamma}(\eta)>0$ is
 used. The Riemann curvature effect indirectly affect the dynamo
 acti on since it can be shown that the Riemann curvature may be
 expressed in terms of the Frenet curvature. Throughout the paper,
 one assumes that the helical structure of the plasma device \cite{12} allows
 us to consider that the Frenet curvature and torsion are constants
 and coincide.
\newpage
Here use is made of the Serret-Frenet holonomic frame \cite{13}
equations, specially useful in the investigation of Riemannian
geometry of plasma flux tubes. Here the Frenet frame is attached
along the magnetic flux tube axis which is endowed with a Frenet
torsion and curvature \cite{13}, which fully determine
topologically, the magnetic filaments, or magnetic streamlines in
the case of the ideal plasma zero resistivity. Dynamical relations
from vector analysis and the theory of curves in the Frenet frame
$(\textbf{t},\textbf{n},\textbf{b})$ are
\begin{equation}
\textbf{t}'=\kappa\textbf{n} \label{1}
\end{equation}
\begin{equation}
\textbf{n}'=-\kappa\textbf{t}+ {\tau}\textbf{b} \label{2}
\end{equation}
\begin{equation}
\textbf{b}'=-{\tau}\textbf{n} \label{3}
\end{equation}
The dynamical evolution equations in terms of time yields
\begin{equation}
\dot{\textbf{t}}=[{\kappa}'\textbf{b}-{\kappa}{\tau}\textbf{n}]
\label{4}
\end{equation}
\begin{equation}
\dot{\textbf{n}}={\kappa}\tau\textbf{t} \label{5}
\end{equation}
\begin{equation}
\dot{\textbf{b}}=-{\kappa}' \textbf{t} \label{6}
\end{equation}
along with the flow derivative
\begin{equation}
\dot{\textbf{t}}={\partial}_{t}\textbf{t}+(\vec{v}.{\nabla})\textbf{t}
\label{7}
\end{equation}
The solenoidal incompressible flow
\begin{equation}
{\nabla}.\textbf{v}=0\label{8}
\end{equation}
The solution shall be given by the magnetic field
\begin{equation}
\textbf{B}=B_{s}(r)\textbf{t}+B_{\theta}(r,s)\textbf{e}_{\theta}\label{9}
\end{equation}
shall be considered here. The magnetic field equations are given by
the solenoidal character of the magnetic field
\begin{equation}
{\nabla}.\textbf{B}=0\label{10}
\end{equation}
where $B_{s}$ is the toroidal component of the magnetic field while
$B_{r}$ and $B_{\theta}$ are respectively radial and poloidal
magnetic fields. The remaining field equation is the self-induction
one
\begin{equation}
{\partial}_{t}\textbf{B}={\nabla}{\times}(\textbf{v}{\times}\textbf{B})+{\eta}{\nabla}^{2}\textbf{B}
\label{11}
\end{equation}
where ${\eta}$ is the magnetic diffusion or resistivity. Though
astrophysical scales ,
${\eta}{\nabla}^{2}\approx{{\eta}{L^{-2}}}\approx{{\eta}{\times}10^{-18}}m^{-2}$
for a solar loop scale length of $10^{8} m$, one notes that the
diffusion effects in plasma LABs are appreciable. Nevertheless for
practical terms one can use the $R_{m}\rightarrow{\infty}$ limit,
which is equivalent to the ${\eta}\rightarrow{0}$. The gradient
operator is given by
\begin{equation}
{\nabla}=\textbf{t}{K}^{-1}{\partial}_{s}+\textbf{e}_{\theta}\frac{1}{r}{\partial}_{\theta}+\textbf{e}_{r}{\partial}_{r}\label{12}
\end{equation}
for the Riemannian line element or metric of flux tube
\begin{equation}
dl^{2}= dr^{2}+r^{2}d{{\theta}_{R}}^{2}+K^{2}(r,s)ds^{2} \label{13}
\end{equation}
where ${\theta}(s):={\theta}_{R}-\int{{\tau}(s)ds}$ and $r_{0}$ is
the constant radius of the constant cross-section flux tube, and
$K(s)=(1-r{\kappa}(s)cos{\theta}(s))$. If the tube is thin factor
$K(s)\approx{1}$. The relations above allows us with a simpler
expression for the Riemannian gradient as
\begin{equation}
{\nabla}=[\textbf{t}{K}^{-1}-\textbf{e}_{\theta}\frac{1}{r}]{\partial}_{s}+\textbf{e}_{r}{\partial}_{r}\label{14}
\end{equation}
This simpler formula allows us to deduce a simpler expression for
the Laplacian
\begin{equation}
{\nabla}^{2}\textbf{B}={{\partial}^{2}}_{r}\textbf{B}+\frac{1}{r}{\partial}_{r}\textbf{B}+\frac{{{\tau}_{0}}^{-2}}{r^{2}}(1+sec^{2}{\theta}){\partial}_{s}
\textbf{B} \label{15}
\end{equation}
where one has consider that helical structure is present where
torsion and curvature of flux tube coincides and are constants and
that ${\partial}_{\theta}=-{{\tau}_{0}}^{-1}{\partial}_{s}$. The
other self-induction equations are
\begin{equation}
{\gamma}+\frac{{B^{0}}_{\theta}}{{B^{0}}_{s}}=\frac{u_{\theta}}{r}tg{\theta}+\frac{{\eta}}{{B^{0}}_{s}}({{\partial}_{r}}^{2}{{B^{0}}_{s}}+\frac{1}{r}{{\partial}_{r}}{{B^{0}}_{s}}
-\frac{{{\tau}_{0}}^{-2}}{r^{3}}) \label{16}
\end{equation}
and
\begin{equation}
[{\gamma}sin{\theta}+{\omega}_{0}cos{\theta}]{B^{0}}_{\theta}=-[\frac{u_{\theta}}{u_{s}}-sin^{2}{\theta}]u_{s}[{B^{0}}_{\theta}+\frac{{B^{0}}_{s}}{cos{\theta}}]
\label{17}
\end{equation}
where here $u_{\theta}={\omega}_{0}r$ , ${\omega}_{0}$ is the
constant rotation of the flow along the Riemannian torus dynamo
\cite{14}. In the next section one shall assume almost the same
value of the plasma laminar dynamo flow, in order to obtain the
topology and geometrical properties of the flow. From the first
equation, with the simplification of weak torsion and curvature one
obtains in the limit when ${\eta}\rightarrow{0}$ one obtains
\begin{equation}
{\gamma}(\eta)=\frac{{B^{0}}_{\theta}}{{B^{0}}_{s}}[{{\kappa}_{0}}^{2}+\frac{u_{\theta}}{r}tg{\theta}+\frac{{\eta}}{{B^{0}}_{s}}({{\partial}_{r}}^{2}{{B^{0}}_{s}}+\frac{1}{r}{{\partial}_{r}}{{B^{0}}_{s}}
-\frac{{{\tau}_{0}}^{-2}}{r^{3}})] \label{18}
\end{equation}
which simplifies to
\begin{equation}
{\gamma}({\eta}\rightarrow{0})=\frac{{B^{0}}_{\theta}}{{B^{0}}_{s}}[1+\frac{u_{\theta}}{r}tg{\theta}]
\label{19}
\end{equation}
which can still be further reduced to
\begin{equation}
{\gamma}({\eta}\rightarrow{0})=\frac{{B^{0}}_{\theta}}{{B^{0}}_{s}}[{{\kappa}_{0}}^{2}+{\omega}_{0}]
\label{20}
\end{equation}
Here one considers that the time dependence of the toroidal and
poloidal fields as
\begin{equation}
B_{s}={{B^{0}}_{s}}e^{{\gamma}t} \label{21}
\end{equation}
\begin{equation}
B_{s}={{B^{0}}_{s}}e^{{\gamma}t} \label{22}
\end{equation}
The other equation is
\begin{equation}
{\omega}_{0}=-\frac{u_{\theta}}{r} \label{23}
\end{equation}
where one has considered the following approximations
\begin{equation}
\frac{{B^{0}}_{\theta}}{{B^{0}}_{s}}<<1 \label{24}
\end{equation}
and also consider that angle ${\theta}<<1$. With this last
expression and the first one , and the a lower bound constraint on
curvature ${\omega}_{0}<<{{\kappa}_{0}}^{2}$ which is suitable for
small Riemannian torus the growth rate ${\gamma}$ reduces to
\begin{equation}
{\gamma}({\eta}\rightarrow{0})=\frac{{B^{0}}_{\theta}}{{B^{0}}_{s}}{{\kappa}_{0}}^{2}
\label{25}
\end{equation}
By considering the Ricca's condition for twisted magnetic flux tube
one obtains
\begin{equation}
\frac{{B^{0}}_{\theta}}{{B^{0}}_{s}}=\frac{{Tw}r}{KL} \label{26}
\end{equation}
where $Tw$ is twist which here one considers to be only proportional
to torsion. Since in twisted torus the torsion coincides with
curvature, one obtains
\begin{equation}
\frac{{B^{0}}_{\theta}}{{B^{0}}_{s}}=\frac{1}{L} \label{27}
\end{equation}
Note that $K\approx{{\kappa}_{0}r}$ and ${Tw}\approx{{\kappa}_{0}}$
yields
\begin{equation}
{\gamma}({\eta}\rightarrow{0})=\frac{{{\kappa}_{0}}^{2}}{L}
\label{28}
\end{equation}
To make an estimate of the growth rate here one considers a small
Riemannian torus of $L=2{\pi}R={\pi}$ and radius of $0.5m$ and this
yields a growth rate of the order ${\gamma}=0.318$. One the notes
that as $Re_{m}$ grows fast, namely from $210$ to $\infty$, the
growth rate of the magnetic field grows fast as well since it goes
from $0.055$ to $0.318$. Thus to conclude one must say that fast
dynamo plasma modes can be excited by curving the Zhang et al
laminar dynamo plasmas in straight cylinders to Riemannian curved
and twisted torus called in plasma physics heliotrons or
stellarators \cite{15}. Let us now consider the computation of the
growth rate by assuming the plasma torus stability, which is given
by considering the quality factor
\begin{equation}
\frac{{B_{0}}^{s}}{{B_{0}}^{s}}\frac{r_{0}}{R}=q \label{29}
\label{29}
\end{equation}
equal to one. This yields
\begin{equation}
\frac{{B_{0}}^{s}}{{B_{0}}^{s}}=\frac{R}{r_{0}} \label{30}
\end{equation}
Let us consider the values of the growth rate obtained from the
self-induction equations above as
\begin{equation}
{\gamma}=\frac{{B^{0}}_{\theta}}{{B^{0}}_{s}}({\omega}_{0}-{{\kappa}_{0}}^{2})
 \label{31}
\end{equation}
\begin{equation}
{\omega}_{0}=\frac{u_{s}}{r}\frac{{B^{0}}_{s}}{{B^{0}}_{\theta}}
\label{32}
\end{equation}
Since the effect of the curvature here is assumed to be small,
${\omega}_{0}>>{{\kappa}_{0}}^{2}$ and the above equations reduce to
\begin{equation}
{\gamma}=\frac{{B^{0}}_{\theta}}{{B^{0}}_{s}}{\omega}_{0}
 \label{33}
\end{equation}
\begin{equation}
{\omega}_{0}=\frac{{\omega}R}{r_{0}}\frac{{B^{0}}_{s}}{{B^{0}}_{\theta}}
\label{34}
\end{equation}
By assuming that the plasma dynamo torus is stable, one may
substitute the equation (\ref{30}) into (\ref{33}) and (\ref{34}) to
yield
\begin{equation}
{\gamma}=\frac{r_{0}}{R}{\omega}_{0}
 \label{35}
\end{equation}
\begin{equation}
{\omega}_{0}=\frac{{\omega}R^{2}}{{r_{0}}^{2}} \label{36}
\end{equation}
Now substitution of (\ref{36}) into (\ref{35}) yields
\begin{equation}
{\gamma}=\frac{{\omega}R}{r_{0}} \label{37}
\end{equation}
By assuming the same ratio $\frac{r}{L}=1$ considered in the laminar
plasmas, one obtains
\begin{equation}
{\gamma}=\frac{{\omega}}{2{\pi}} \label{38}
\end{equation}
By considering now the Lau-Finn \cite{16} oscillation of the chaotic
dynamo flow as ${\omega}=\frac{2{\pi}}{6}$ this expression reduces
to ${\gamma}=\frac{{\omega}}{2{\pi}}=1/6\approx{0.16}$ which is even
higher than the chaotic fast dynamo obtained by Lau-Finn result
which has a growth rate of ${\gamma}=0.077$ for $Re_{m}={\infty}$.
They also obtain for $Re_{m}={1000}$ a ${\gamma}=0.076$, showing a
similar behaviour that was obtained previously in our paper here,
that to dramatically increase the Reynolds number after some limit
does not grow appreciable the value of ${\gamma}$. From expression
(\ref{38}) one obtains ${\gamma}=0.115{{\omega}}$ which is a smaller
value than the one obtained from Vainshtein et al \cite{17} of
${\gamma}=0.485{{\omega}}$ of the growth rate in terms of the
toroidal vorticity. Growth rates as high as $3$ have been found in
Perm liquid sodium dynamo torus, as has been computed by Dobler et
al \cite{18}.

\newpage
\textbf{Acknowledgements:} \newline I thank the late Vladimir Tsypin
for helpful discussions on the subject of plasma physics and
Riemannian geometry. Special thanks go to G\"{u}nther R\"{u}diger
and Dymitry Sokoloff for helpful discussions on Perm torus dynamo
and Riemannian geometry in dynamo theory. Financial supports from
Universidade do Estado do Rio de Janeiro (UERJ) and CNPq (Brazilian
Ministry of Science and Technology) are highly appreciated.

  \end{document}